\documentclass[12pt,preprint]{emulateapj}
\usepackage[]{natbib}
\usepackage[]{rotating}

\shortauthors{R. E. Louis et al.}

\begin{document}

\title{Properties of Umbral Dots from Stray Light Corrected {\em Hinode} Filtergrams\\}

\author{Rohan E. Louis\altaffilmark{1}$^{\textrm{,}}$\altaffilmark{2}, Shibu K. Mathew\altaffilmark{2}, 
Luis R. Bellot Rubio\altaffilmark{3}, Kiyoshi Ichimoto\altaffilmark{4},
B. Ravindra\altaffilmark{5} and A. Raja Bayanna\altaffilmark{2}}
\altaffiltext{1}{Presently at Leibniz-Institut f$\ddot{\textrm{u}}$r Astrophysik Potsdam (AIP),
                  An der Sternwarte 16,
	       	  14482 Potsdam, Germany}

\altaffiltext{2}{Udaipur Solar Observatory, Physical Research Laboratory,
                  Dewali, Badi Road, Udaipur,
	       	  Rajasthan - 313004, India}

\altaffiltext{3}{Instituto de Astrof\'{\i}sica de Andaluc\'{\i}a (CSIC),
                     Apartado de Correos 3004,
                     18080 Granada, Spain}

\altaffiltext{4}{Kwasan and Hida Observatories,
		  Kyoto University, Yamashina-ku, 
		  Kyoto 607-8417, Japan}

\altaffiltext{5}{Indian Institute of Astrophysics, 
		  II Block, Koramangla,
		  Bangalore - 560 034, India}

\email{rlouis@aip.de}

\begin{abstract}
High resolution blue continuum filtergrams from {\em Hinode} are 
employed to study the umbral fine structure of a regular unipolar 
sunspot. The removal of scattered light from the 
images increases the $rms$ contrast
by a factor of 1.45 on average. Improvement in image 
contrast renders identification of short filamentary 
structures resembling penumbrae that are well separated from the 
umbra-penumbra boundary and comprise bright filaments/grains flanking 
dark filaments. Such fine structures were recently detected from
ground based telescopes and have now been observed with {\em Hinode}. A multi-level 
tracking algorithm was used to identify umbral dots in both the 
uncorrected and corrected images and to track them in 
time. The distribution of the values describing the photometric and 
geometric properties of umbral dots are more easily affected by the presence of stray 
light while it is less severe in the case of kinematic properties. 
Statistically, umbral dots exhibit a peak intensity, effective diameter, 
lifetime, horizontal speed and a trajectory length of  0.29$I_{\textrm{\tiny{QS}}}$, 
272 km, 8.4 min, 0.45 km~s$^{-1}$ and 221 km respectively. 
The 2 hr 20 min time sequence depicts several locations where umbral dots 
tend to appear and disappear repeatedly with various time intervals.
The correction for scattered light in the {\em Hinode} filtergrams 
facilitates photometry of umbral fine structure which can be related
to results obtained from larger telescopes and numerical simulations.
\end{abstract}



\section{Introduction}
\label{intro}
The dark umbral background is populated by small, bright 
features called umbral dots (UDs). The size of UDs ranges from
0\farcs8 to about 0\farcs2 based on previous studies 
by \citet{Sobotka1997a} and \citet{Tritschler2002}, while 
a recent work by \citet{Rth2008b} shows that the size 
distribution of UD diameters is a maximum around 0\farcs3 or 225 km, suggesting 
that most of the UDs are spatially resolved. \citet{Sobotka1997a} 
also observed that the larger, long-lived UDs are seen in regions 
of enhanced umbral background intensity. The darkest parts of the 
umbral core, referred to as dark nuclei, are often devoid of UDs. Based 
on their relative location, UDs can be classified as ``central'' and 
``peripheral''. While the former are seen in the inner regions of the 
umbra, the latter dominate the umbra-penumbra boundary. Peripheral 
UDs are usually brighter than the central ones. The intensity of UDs 
ranges from about 0.2 to 0.7 times the normal photospheric 
intensity at visible wavelengths. The typical speeds of UDs are 
$\approx$ 400 m~s$^{-1}$ \citep{Sobotka1997b,Kitai2007,Rth2008b,Sobotka2009b}.
Most mobile UDs emerge near the umbra-penumbra boundary and move towards 
the center of the umbra \citep{Rth2008b} with speeds of 700 m~s$^{-1}$. 
UDs do not have a typical lifetime, with values ranging from 10 min 
to 2.5 min \citep{Sobotka1997a,Rth2008b,Hamedivafa2011,Watanabe2009b}. 
The spread in the values arises from the manner in which UDs can be grouped based on
their size, lifetime and spatial location.

\citet{Parker1979} proposed that UDs are manifestations of hot non-magnetized 
plasma pushing its way in the gappy umbral field. While the detection 
of such weak fields in the umbra remains elusive, observations 
indicate a reduction of 300-500 G in UDs \citep{Schmidt1994,Hector2004} 
with the contrasted ones residing in locations where the magnetic 
field is $\approx$ 2000 G and is inclined more than 
30$^{\circ}$ \citep{Watanabe2009b}. Central 
and peripheral UDs exhibit an enhancement in temperature of 550 K and 
570 K respectively \citep{Rth2008a}. The measurements of Doppler 
velocities in UDs show that peripheral UDs have an upflow 
of $\approx$ 0.4-0.8 km~s$^{-1}$ \citep{Rimmele2004,Rth2008a,Sobotka2009a}. 
Central UDs on the other hand exhibit very weak 
downflows \citep{Sobotka2009b} while \citet{Hartkorn2003} detected 
downflows of upto 0.3 km~s$^{-1}$. \citet{Ortiz2010} 
reported downflows at the edge of UDs measuring 400 to 1000 m~s$^{-1}$ at a 
spatial resolution of 0\farcs14. 3D MHD simulations of \citet{Schussler2006}, which 
model UDs as narrow upflowing plumes in regions of intense magnetic fields, 
predicted a central dark lane in UDs which was subsequently verified
from high resolution ground based observations 
\citep{Rimmele2008,Sobotka2009b,Ortiz2010} as well as 
from space \citep{Bharti2007}. 
Recent observations from the 1.6 m NST indicate
that UDs are not perfectly circular but possess a mean eccentricity of 0.74
in the photosphere \citep{Kilcik2011}.

The extent to which UDs can be resolved depends on the spatial resolution 
which is presently 0\farcs14 \citep{Sobotka2009b,Ortiz2010} for 
a 1 m ground based telescope at 450 nm. This has been made possible with the aid of 
adaptive optics and post-processing techniques to minimize the contribution 
from ``seeing''. On the other hand, {\em Hinode} \citep{Kosugi2007} with a 
50 cm aperture in space has a resolution of 0\farcs23 at the same wavelength
which can operate for long periods in the absence of the Earth's atmospheric turbulence.
Furthermore, since UDs are present in the darkest regions of sunspots, 
contamination by stray light can strongly influence photometric investigation
of these structures. The motivation of this paper is to study the influence of
stray light on the properties of UDs using {\em Hinode} data and how its 
removal compares with existing high resolution ground-based observations 
as well as numerical simulations.
This exercise has been 
done taking into account the trade off between marginally coarser spatial 
resolution and the uninterrupted ``seeing-free'' time sequence between {\em Hinode} 
and its 1 m ground based counterpart. The rest of the paper is 
organized as follows. The observations are described in Section~\ref{data} 
and the results are presented in Section~\ref{res}. In Section~\ref{summary} we
summarize our findings and discuss their implications.

\begin{figure}[!h]
\hspace{-15pt}
\centerline{
\includegraphics[angle=90,width = 0.51\textwidth]{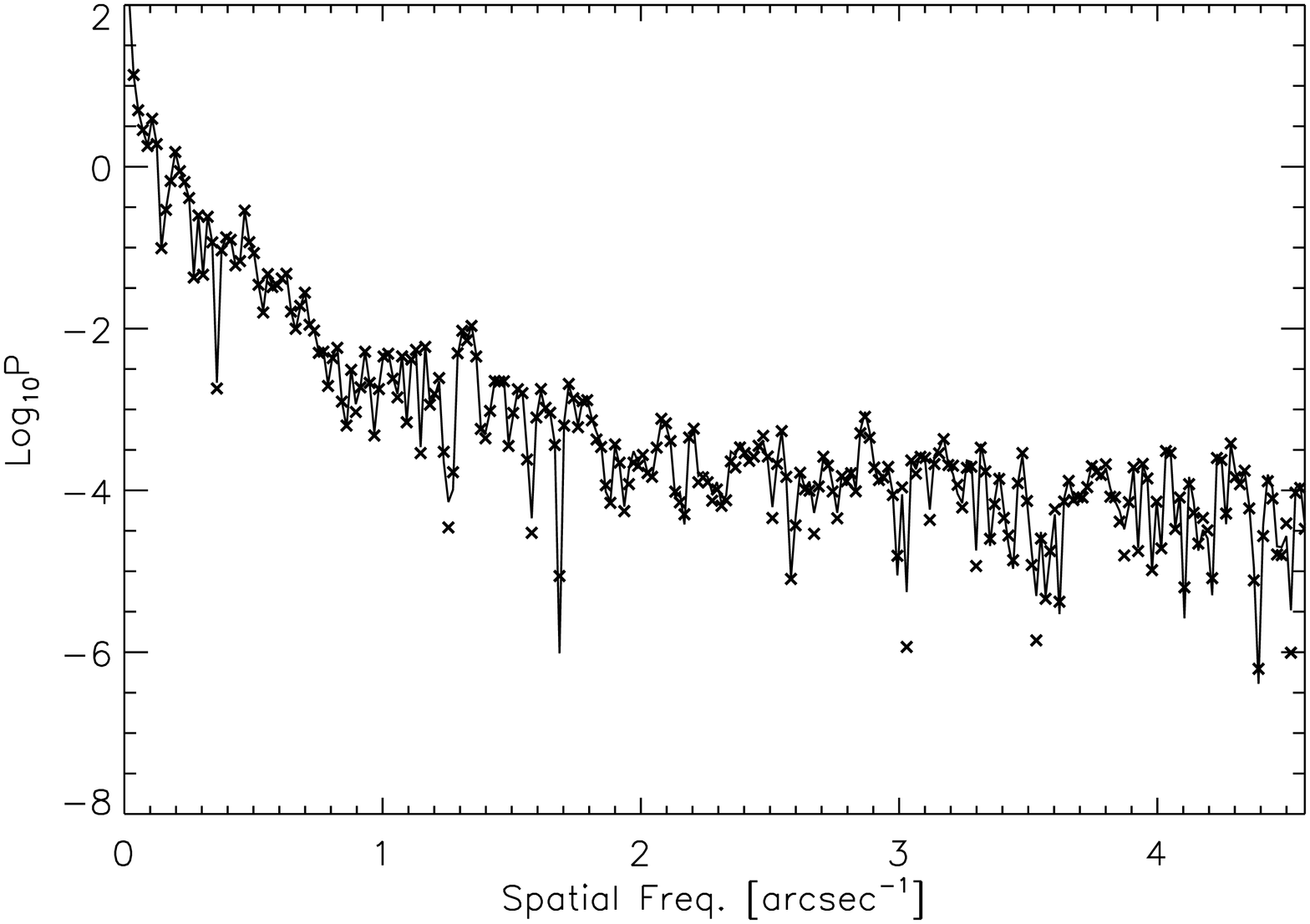}
}
\vspace{-10pt}
\centerline{
\includegraphics[angle=90,width = 0.51\textwidth]{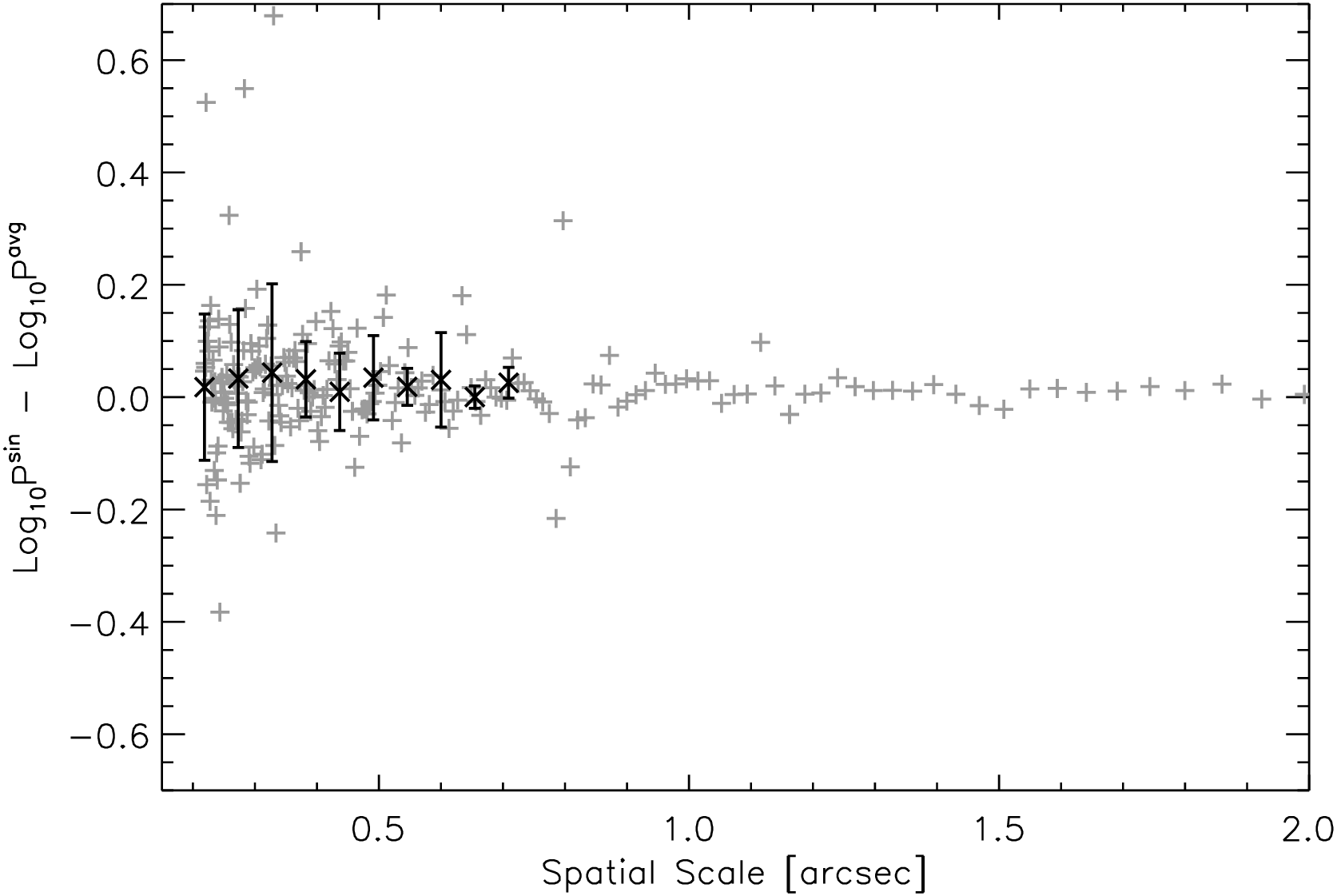}
}
\vspace{-10pt}
\caption{{\bf{Top: }}Azimuthally averaged power spectra of single ({\em solid} line) and 
average of 4 blue continuum filtergrams ({\em cross} symbol). {\bf{Bottom: }}Difference 
in power spectra of a single and average filtergram as a function of spatial scale 
({\em grey plus} symbols). The {\em black cross} symbols represent average values of the 
difference in bins of 0\farcs05 with the vertical bars denoting the {\em rms} value.}
\label{power_spec}
\end{figure}

\section{Observations and Data Processing}
\label{data}
We utilize high resolution blue continuum filtergrams of the 
sunspot in NOAA AR 10944 acquired by the BFI 
(Broadband Filter Imager) of the SOT \citep[Solar Optical 
Telescope;][]{Tsuneta2008}, on board {\em Hinode} from 00:14 - 02:34 UT
on March 1, 2007. The sunspot was located very close to 
disc center (N0.7W4) at a heliocentric angle of $\Theta = $ 4$^\circ$.
The 1024$\times$512 filtergrams had a pixel sampling of 0\farcs054 and were taken 
at a cadence of 6 s with an exposure time of 102 ms. Initial processing of 
Level-0 data to Level-1 included dark correction, flat fielding and
removal of bad pixels, and was carried out using the ``{\em fg\_prep}'' 
routine in SolarSoft. The images were subsequently co-aligned using a 2D cross
correlation routine.

In order to reduce the noise in the dark regions of the umbra, 
four successive filtergrams were added to yield a sequence of 326 images 
with a cadence of $\sim$ 26 s. The averaging was carried out to keep the 
temporal resolution identical to the analysis of \citet{Watanabe2009b} 
who had previously studied the same AR with the above data set. 
Applying the running average to the filtergrams leads to a reduction of power especially in 
the high spatial frequency domain as shown in the top panel of 
Figure~\ref{power_spec} which depicts the azimuthally averaged power spectrum.
The bottom panel of Figure~\ref{power_spec} 
illustrates the difference in power between the single and mean image 
as a function of the spatial scale.
Since the power is proportional to the square of the intensity, 
the relative intensity difference between the images varies from 
2\% to 5\% for spatial scales of 0\farcs2 and 0\farcs5, respectively. 
This implies that averaging successive filtergrams should not significantly affect 
the detection of small scale structures within the umbra. 
We estimate the intensity fluctuations in the umbra to be 1.5\% of the 
quiet Sun (QS) intensity. This value corresponds to three times the standard 
deviation of a small region in the umbral dark core.
Averaging the filtergrams reduces the noise by $\approx$6\%.

{\em Insrumental Stray Light: }The blue continuum filtergrams were corrected 
for instrumental stray light using the PSF (Point Spread Function) described 
by \citet{Shibu2009}. This PSF was derived from transit observations of Mercury 
on November 8, 2006. \citet{Shibu2009} showed that the removal of stray light 
renders an improvement in the contrast of bright points in the quiet Sun by a 
factor of 2.4-2.75 at 430 nm. As the transit observations were taken close to 
disc center, the same PSF was utilized for removing stray light in the blue 
continuum filtergrams. The PSF is a weighted linear combination of 4 
Gaussians \citep[see Table 1 of][]{Shibu2009}. The deconvolution is carried 
out using an IDL maximum likelihood routine\footnote{\tt Called Max\_Likelihood.pro, 
from the AstroLib package} \citep{Richardson1972,Lucy1974}. 
The method uses the instrument PSF to iteratively update the current estimate 
of the image by the product of the previous deconvolution and the correlation 
between re-convolution of the subsequent image and the PSF. 
The algorithm can be expressed as follows:
\begin{equation}
I = O\ast P \\
\end{equation}
The Image ($I$) is a result of the convolution between the Object ($O$) and the Point
Spread Function $P$.  Given $I$ and $P$, the most likely $O$ can be iteratively determined
as:
\begin{equation}
O^{t + 1} = O^t \cdot \left (\frac{I}{C} \ast \textrm{conj} P \right)\\
\end{equation}
where $C = O^t\ast P$ under the assumption of Poisson 
statistics. Here $t$ refers to the iteration cycle.

\begin{figure}[!h]
\centerline{
\hspace{0pt}
\includegraphics[angle=90,width = 0.5\textwidth]{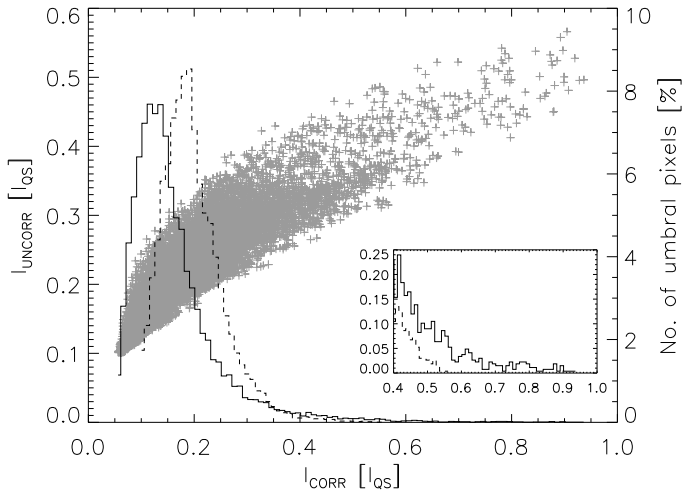}
}
\caption{Change in umbral intensity after removal of stray light. Displayed is a scatter plot of uncorrected 
(${\textrm{I}}_{\textrm{\tiny{uncorr}}}$) and corrected (${\textrm{I}}_{\textrm{\tiny{corr}}}$) 
blue continuum intensity in the umbra. Note that the axes are scaled differently. The {\em solid} 
and {\em dashed} lines refer to the intensity histograms of the corrected and uncorrected 
filtergrams respectively with a binsize of 0.01$I_{\textrm{\tiny{QS}}}$. A magnified version of 
the trailing halves of the histograms is shown in the inset.}
\label{umb_sta}
\end{figure}

\begin{figure}[!h]
\centerline{
\includegraphics[angle=90,width = 0.547\textwidth]{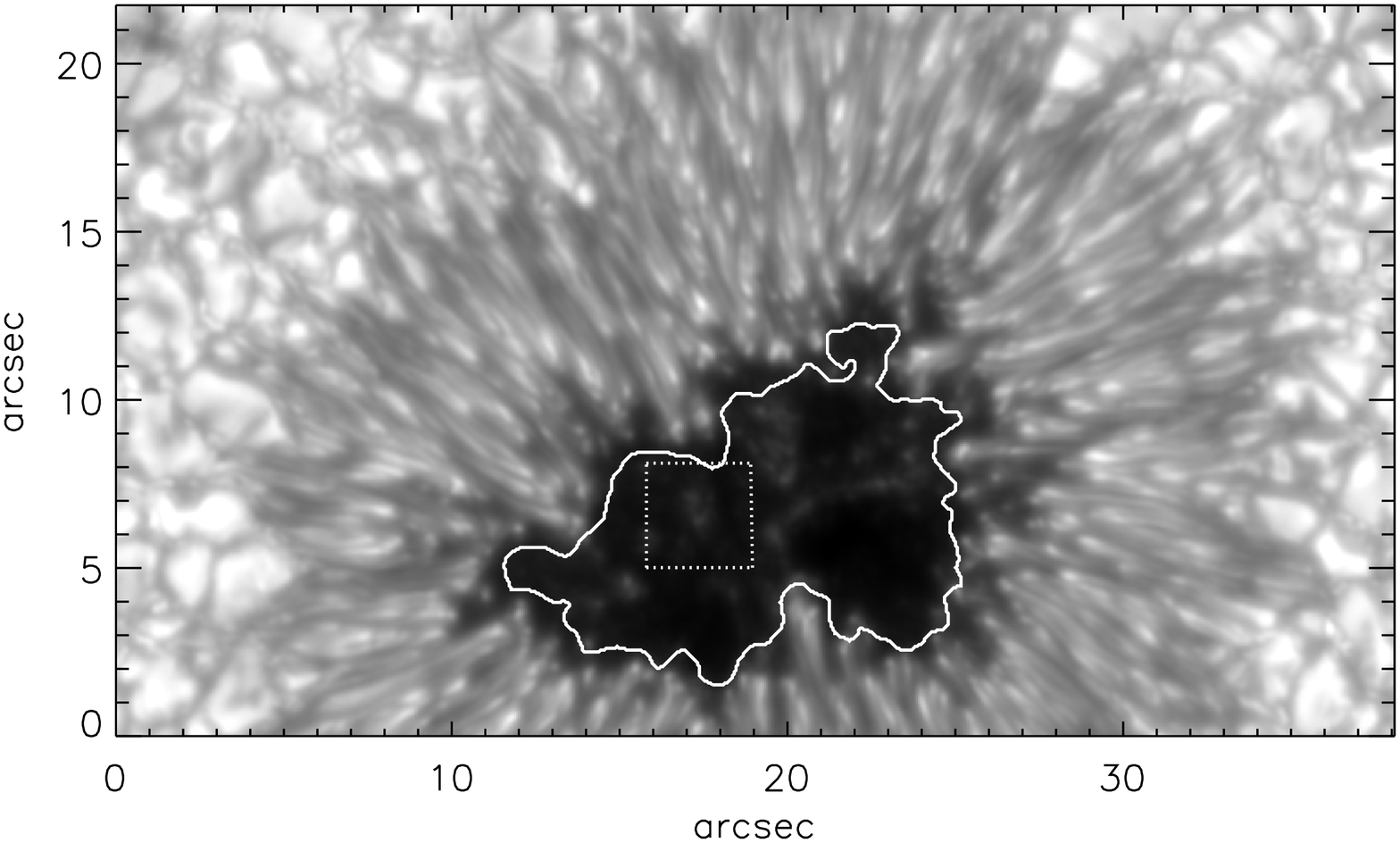}
}
\vspace{-23pt}
\centerline{
\includegraphics[angle=0,width = 0.547\textwidth]{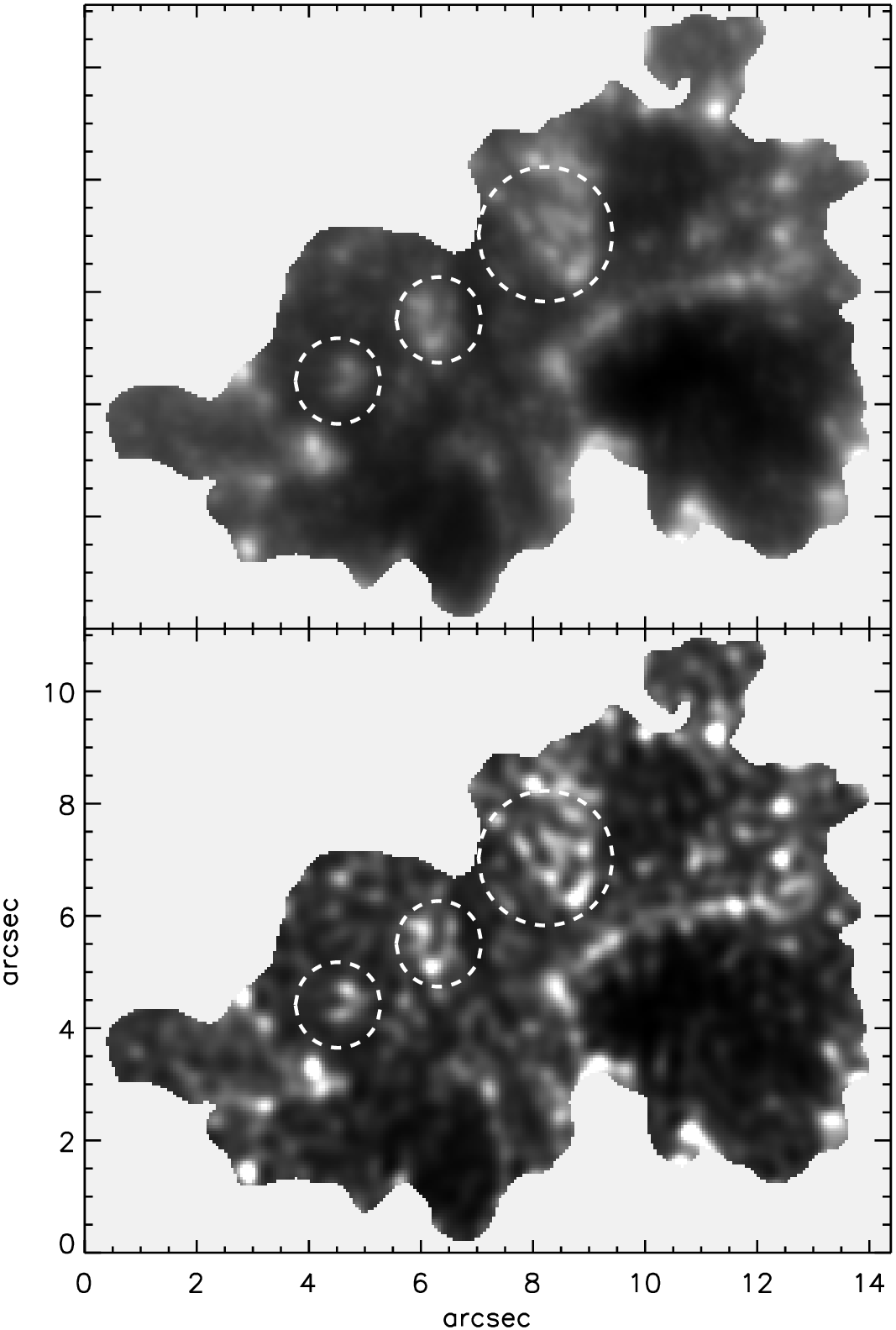}
}
\vspace{-20pt}
\caption{Improvement in image contrast after removal of stray light. {\bf{Top: }}Sunspot in 
NOAA AR 10944. The white contour refers to the umbral area chosen for analysis. The 
{\em dotted white} box representes a selected field-of-view shown in Figure~\ref{filam}. 
{\bf{Middle: }}Magnified view of umbra before stray light correction. {\bf{Bottom: }}Stray 
light corrected image. The images in the middle and bottom panels have been scaled identically. 
The {\em white dashed} circles enclose short filaments that are described in Section~\ref{fine}.}
\label{umb}
\end{figure}

The corrected and 
uncorrected sequences were subsequently normalized to the QS intensity.
Figure~\ref{umb_sta} shows a typical scatter plot of intensities in a 
corrected filtergram and the corresponding uncorrected image. Following 
stray light correction, the minimum umbral 
intensity corresponding to the dark umbral core, reduces from 0.1$I_{\textrm{\tiny{QS}}}$ 
to 0.05$I_{\textrm{\tiny{QS}}}$ which is above the estimated noise level in
the the filtergrams. In addition, the fraction of pixels in the umbra having 
an intensity less than 0.3$I_{\textrm{\tiny{QS}}}$ in both sets of images 
is $\approx$ 93\%. 
The removal of stray light decreases the mean umbral intensity from 0.202$I_{\textrm{\tiny{QS}}}$ to 
0.167$I_{\textrm{\tiny{QS}}}$ while the $rms$ contrast increases from 0.059 to 0.086 which is a factor 
of 1.45.
The middle and bottom panels of Figure~\ref{umb} show the
uncorrected and corrected image respectively.
The umbral region is extracted from the sunspot shown as the contoured 
region (top panel of Figure~\ref{umb}).
The stray light corrected image exhibits various
fine scale features which are described in the following 
section. 
Hereafter `uncorrected' and `corrected' will be
referred to as UN and CR respectively \footnote{The movie of the uncorrected 
and corrected images is available at {\em www.prl.res.in/$\sim$eugene/movie-umbral-dots.wmv}}.

\section{Results}
\label{res}

\subsection{Umbral Fine Structure}
\label{fine}
{(a) \em Short Filaments: }Figure~\ref{umb} depicts three examples of 
short filaments (shown in {\em dashed circles}) near the periphery of the umbra 
that closely resemble penumbral filaments. These short structures
have varying lengths with a dark lane and two adjacent brightenings, and 
a bead-like brightening at the tip of the filament facing
the umbra. The width of the filaments shown in the figure range from 165 - 200 
km where the latter can be considered an upper limit for
these structures. These dark filaments are similar to the ones reported by
\citet{Rimmele2008,Sobotka2009b} and in numerical simulations 
\citep{Rempel2009} . However, to the best of 
our knowledge this is the first time that such filaments have been 
seen in {\em Hinode} observations. Their morphology is different from 
the traditional dark lane associated with UDs as observed by \citet{Rimmele2008,Sobotka2009b} 
and \citet{Ortiz2010}. The filament indicated by the largest circle 
in the bottom panel of Figure~\ref{umb} partially obscures a neighboring 
filament to its right. 

\begin{figure}[!h]
\centerline{
\includegraphics[angle=90,width = 0.5\textwidth]{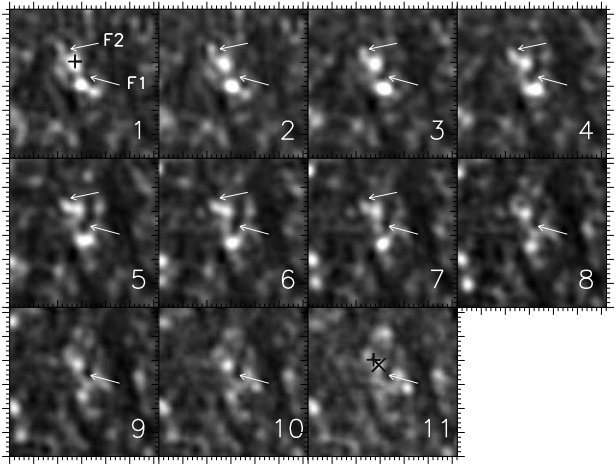}
}
\vspace{-5pt}
\caption{Temporal evolution of dark filaments and umbral dots. The images correspond to the 
field-of-view indicated by the {\em dotted white} box in Figure~\ref{umb}. The frames have a 
cadence of 100s and show an UD traveling from the end of one dark filament F2 to the flank 
of another filament F1. The {\em black plus} and {\em cross} symbols denote the initial and final position
of the UD in the time sequence respectively. Each major tickmark corresponds to 0\farcs5.}
\label{filam}
\end{figure}

Figure~\ref{filam} depicts the temporal evolution of a set of dark filaments
and bright UDs. One can identify at least two main filaments 
labeled F1 and F2. The time separation between individual frames is 100 s.
The lifetime of F1 and F2 is estimated to be 18 and 10 min respectively. These
values can be regarded as lower limits since both of them were present 
from the start of the sequence. The {\em black plus} symbol marks the position
of a bright grain/UD which is seen during the chosen sequence. The UD starts at
the end of F2 and ends adjacent to F1 on its left. The total displacement of the UD
from frame 1 to 11 (cross symbol) is $\sim$0\farcs15 for the 18 min sequence. 
In general, these localized brightenings/UDs move from the tip/end of one
filament onto the flank of another. This motion is continued till they can no 
longer be distinguished from the background. The dark filaments usually
maintain their form during the motion of the UDs after which they can either
diffuse or break up into even smaller dark segments.
\begin{figure}[!h]
\centerline{
\includegraphics[angle=90,width = 0.5\textwidth]{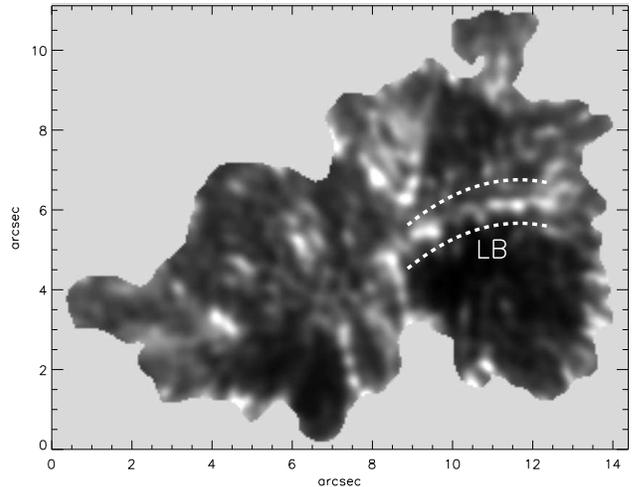}
}
\caption{Average image constructed from 120 frames in the stray light corrected sequence with 
the white {\em dashed} lines indicating a faint light bridge.}
\label{lb}
\end{figure}

{(b) \em Light Bridges: }
In addition to UDs, sunspot umbrae often exhibit light bridges (LBs). 
These can be broadly classified into two categories, namely - strong 
and faint \citep{Sobotka1993,Sobotka1994}. While the former split the 
umbra into individual cores and represent an abrupt change in the 
umbral morphology, the latter are usually less than 1$''$
in width and are composed of a chain of UDs.  
The top panel of Figure~\ref{umb} shows the umbra to be devoid of 
any large scale structuring, but Figure~\ref{lb} indicates a faint 
LB near the right-hand side of the umbra, that is nearly horizontally 
orientated above an umbral dark core. The image is an average of 
120 filtergrams covering a 50 min duration. 
The LB consists of several bright grains resembling UDs whose width 
is $\approx$0\farcs23 and is close to the resolution limit of {\em Hinode}. 
The average grain spacing on the LB is estimated to be 0\farcs25 while 
the length of the LB is nearly 3.8$''$.

\subsection{Properties of UDs}
\label{prop}
This section describes the physical properties of UDs which were 
determined from the time sequence of the uncorrected as well as 
corrected blue continuum image sequences. Identifying and tracking of
UDs involves the following steps--

\begin{figure}[!h]
\centerline{
\includegraphics[angle=90,width = 0.5\textwidth]{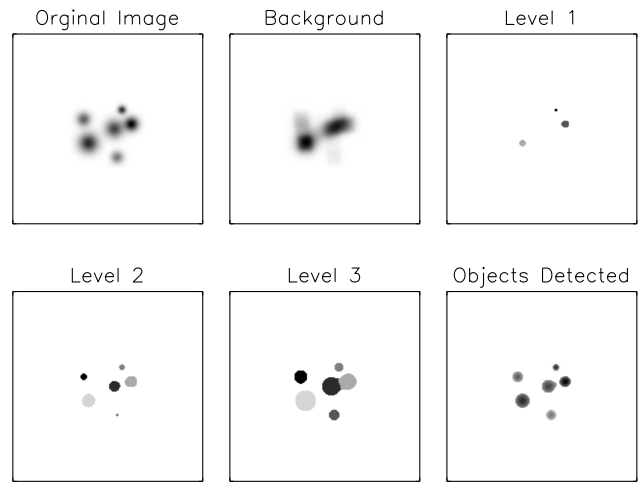}
\vspace{-70pt}
}
\caption{Depiction of MLT algorithm for 3 intensity levels. The original image comprises of an 
arbitrary distribution of intensities. The maximum and minimum intensity range is split into 3 
intermediate levels and the objects are sequentially tagged from the highest to the lowest level. 
The object boundary is determined using the background image from which the features can be 
isolated as shown in the bottom right panel.}
\vspace{10pt}
\label{mlt}
\end{figure}

\begin{enumerate}
\item{}{\em Defining the umbra-penumbra boundary}: The sequence of corrected
images are added to obtain a mean image. After smoothing the mean image using
a 11$\times$11 pixel boxcar, the umbra-penumbra boundary is defined  
by a single continuous contour corresponding to an intensity of 0.3$I_{\textrm{\tiny{QS}}}$.
This contour is used to construct a binary mask which allowed us to extract 
the umbral region from individual images.
\item{}The identification of UDs was carried out using a 2D MLT 
\citep[Multi Level Tracking;][]{Bovelet2001} algorithm  that has 
been described by \citet{Rth2008b} for detecting umbral dots. The algorithm
identifies objects at different intensity levels starting from 
the highest level and tagging them uniquely while progressing to lower intensities, 
till the minimum level is reached. The number of objects (NOs) detected depends 
on the number of levels (NLs) defined. The latter was chosen by 
implementing the algorithm for different levels and counting the total 
NOs detected. For both the UN and CR images, the NOs increase exponentially 
with the NLs. A non-linear least square fit provided an optimal value of 33 
and 25 intensity levels for the UN and CR images respectively. The above 
values correspond to the knee of the best fit. 
\item{}{\em Defining the background image}: The thin plate spline 
technique \citep{Barrodale1993} was employed to construct the background 
umbral image, i.e. the intensity distribution in the absence of UDs. 
Each UD is defined by the intensity contour corresponding to
$(I_{\textrm{\tiny{max}}} + I_{\textrm{\tiny{bg}}})/2$, where 
$I_{\textrm{\tiny{max}}}$ and $I_{\textrm{\tiny{bg}}}$ refer to the 
maximum/peak and background intensity respectively. Figure~\ref{mlt} depicts the 
functionality of the algorithm based on 3 levels for an arbitrary distribution of features.
For each UD, the following quantities are determined: $I_{\textrm{\tiny{max}}}$, 
$I_{\textrm{\tiny{bg}}}$, $I_{\textrm{\tiny{mean}}}$ and 
$D_{\textrm{\tiny{eff}}}$. The effective diameter $D_{\textrm{\tiny{eff}}}$ 
expressed in km, is calculated as $\sqrt{4A/\pi}$ where $A$ is the 
total number of pixels. In addition, the spatial location of the 
maximum intensity ($X_{\textrm{\tiny{p}}}$, $Y_{\textrm{\tiny{p}}}$) is 
also noted. This information is required to track the UDs in the 
image sequence. Before saving the above information for each UD, the 
routine also verifies if the UDs are separated from the umbra-penumbra 
boundary. Only those features which are at least 2 pixels inward from the edge 
of the umbral mask are saved and considered for analysis. The fraction of 
objects that did not get filtered using the above criteria is less than 5\%.
\item{}{\em Tracking the UDs in time}: Each UD is identified in the 
successive frame if $X^i_{\textrm{\tiny{p}}}$, $Y^i_{\textrm{\tiny{p}}}$
and $X^{i+1}_{\textrm{\tiny{p}}}$, $Y^{i+1}_{\textrm{\tiny{p}}}$ are at the most
1 pixel apart (where $i$ refers to the frame index). The condition of 
1 pixel separation is to ensure that the horizontal 
speed of the UD does not exceed 1.5 km~s$^{-1}$ which
is obtained from the spatial sampling of 0\farcs05 and a cadence of 26 s. 
If an UD cannot be identified in the current frame the tracking is extended to 
2 successive frames. If this also fails, the tracking is terminated.
\item{}Once tracking is completed, the following additional properties are determined 
from the UD trajectories: lifetime ($T$), horizontal speed ($V$), birth-death
distance ($L_{\textrm{\tiny{bd}}}$) and trajectory length ($L_{\textrm{\tiny{tj}}}$).
The effective diameter as well as the mean, maximum and ratio of maximum-to-background 
intensities are averaged over the trajectory of the UD. The horizontal velocity is 
calculated as the ratio of the trajectory length and the lifetime.
\item{}Categorizing UDs into peripheral and central is carried out using a second 
boundary that lies $\approx$0\farcs8 inward from the original umbral mask. UDs which 
originate in between these two contours are labeled as peripheral and the rest 
as central.
\end{enumerate}

\begin{figure}[!h]
\centerline{
\hspace{15pt}
\includegraphics[angle=90,width = 0.53\textwidth]{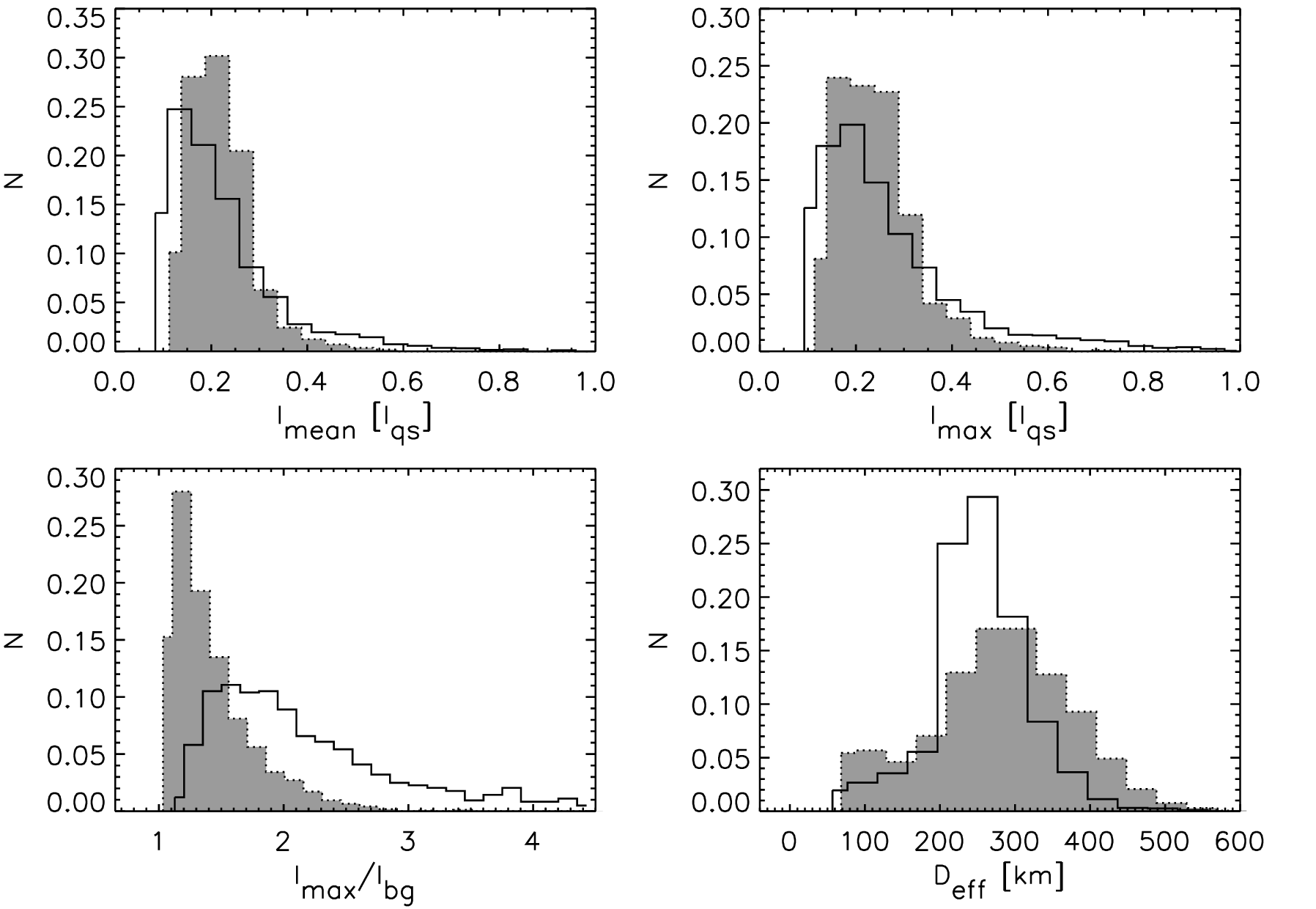}
}
\centerline{
\hspace{15pt}
\includegraphics[angle=90,width = 0.53\textwidth]{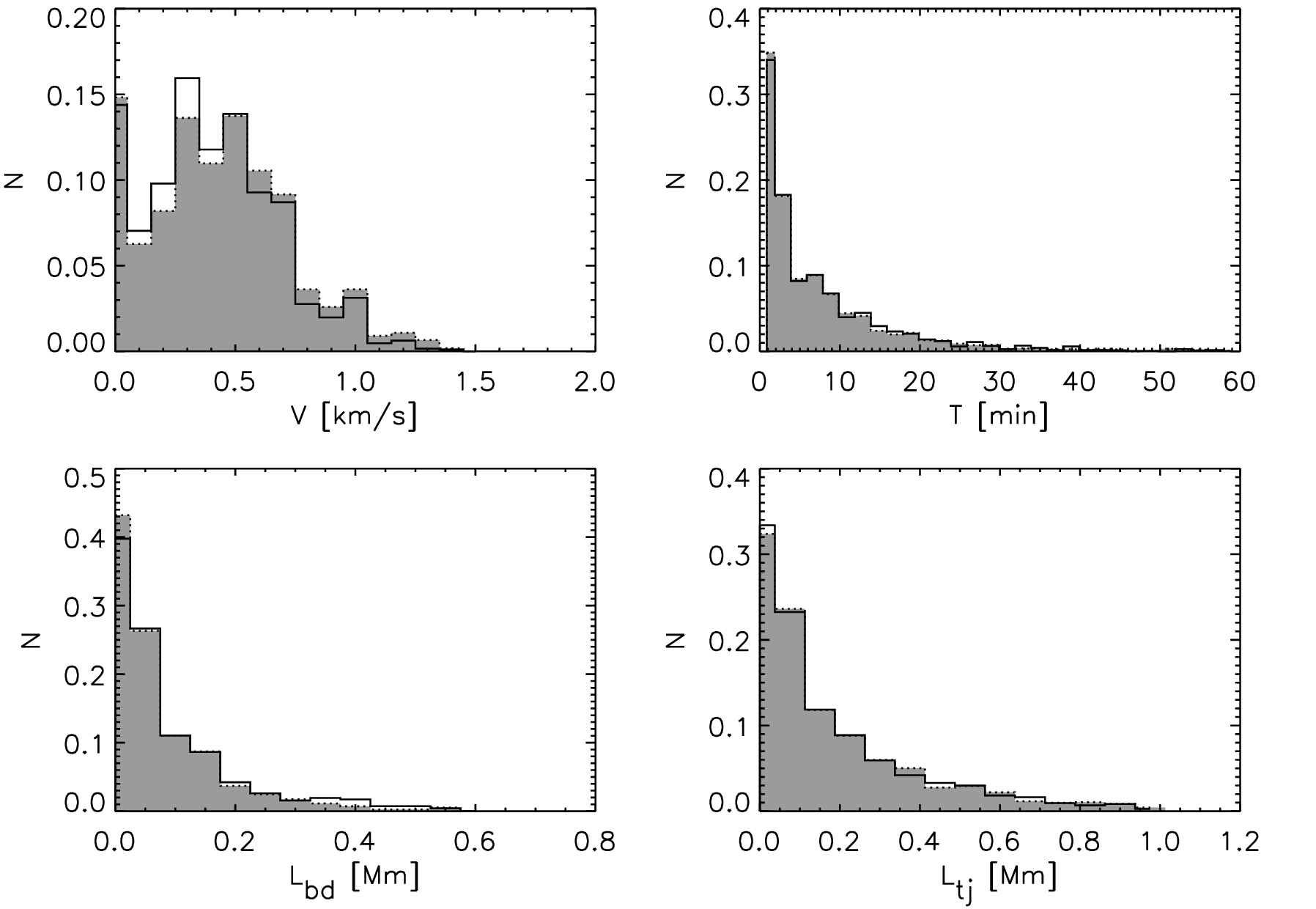}
}
\caption{Histogram of various UD properties for both time sequences. Mean 
Intensity - $I_{\textrm{\tiny{mean}}}$, Maximum Intensity  - $I_{\textrm{\tiny{max}}}$, 
maximum-to-background intensity ratio - $I_{\textrm{\tiny{max}}}$/$I_{\textrm{\tiny{bg}}}$, 
effective Diameter - $D_{\textrm{\tiny{eff}}}$, horizontal speed - $V$, lifetime - $T$, 
birth-death distance - $L_{\textrm{\tiny{bd}}}$ and trajectory length - $L_{\textrm{\tiny{tj}}}$. 
Bin sizes are 0.05$I_{\textrm{\tiny{QS}}}$, 0.05$I_{\textrm{\tiny{QS}}}$, 0.15, 40 km, 
100 m~s$^{-1}$, 2 min, 50 km, 75 km respectively. The grey shaded and unshaded histograms 
corresponds to the uncorrected and corrected time sequences respectively. The $y$-axis represents 
the fraction of UDs in each bin.}
\label{histo}
\end{figure}

Figure~\ref{histo} shows the histogram of the various physical properties of all
UDs (central and peripheral) whose lifetime exceeds 50 s (2 frames). These UDs constitute
$\sim$95\% of the total number of objects detected and tracked in both time sequences. 
The histograms have been normalized to unity for comparison.
The bin size of the quantities corresponding to the UN and CR data is identical.
We now turn to the quantitative differences between the physical properties
of UDs derived from the uncorrected and corrected sets and whose mean values are
summarized in Table~\ref{tab}. \\

\begin{table*}[!ht]
\begin{center}
\caption{Mean value of physical properties of UDs averaged over their trajectories along 
with their $rms$ value. The numbers in the parantheses denote the number of UDs in that group.}
\label{tab}
\begin{tabular}{|c|cc|cc|cc|}
\hline
 & \multicolumn{2}{c|}{All UDs} & \multicolumn{2}{c|}{Central UDs} & \multicolumn{2}{c|}{Peripheral UDs}\tabularnewline \cline{2-7} 
& & & & & & \\
Parameter & No Stray Corr. & With Stray Corr.  & No Stray  Corr. & With Stray Corr. &  No Stray Corr. & With Stray Corr. \\
 & (1690) & (1949) & (953) & (1096) & (737) & (853) \\
\hline \hline
& & & & & & \\
$I_{\textrm{\tiny{mean}}}$ ($I_{\textrm{\tiny{QS}}}$) & 0.24$\pm$0.07 & 0.24$\pm$0.13 & 0.21$\pm$0.05 & 0.21$\pm$0.09 & 0.27$\pm$0.08 & 0.28$\pm$0.16 \\ 
$I_{\textrm{\tiny{max}}}$ ($I_{\textrm{\tiny{QS}}}$) & 0.26$\pm$0.09 & 0.29$\pm$0.17 & 0.23$\pm$0.06 & 0.25$\pm$0.12 & 0.30$\pm$0.10 & 0.34$\pm$0.21 \\ 
$I_{\textrm{\tiny{max}}}$/$I_{\textrm{\tiny{bg}}}$ & 1.48$\pm$0.34 & 2.50$\pm$1.26 & 1.43$\pm$0.28 & 2.33$\pm$0.91 & 1.54$\pm$0.39 & 2.72$\pm$1.57 \\ 
$D_{\textrm{\tiny{eff}}}$ (km) & 295$\pm$101 & 272$\pm$68.0  & 337$\pm$81.0  & 298$\pm$54.0  & 241$\pm$98.0  & 239$\pm$69.0 \\ 
$V$ (km~s$^{-1}$) & 0.49$\pm$0.35 & 0.45$\pm$0.32 & 0.51$\pm$0.36 & 0.46$\pm$0.32 & 0.46$\pm$0.33 & 0.44$\pm$0.32 \\ 
$T$ (min) & 8.80$\pm$11.9 & 8.40$\pm$10.5 & 7.10$\pm$8.70 & 8.90$\pm$11.7 & 10.9$\pm$14.7 & 7.80$\pm$8.70 \\ 
$L_{\textrm{\tiny{bd}}}$ (km) & 94.$\pm$120   & 112$\pm$144   & 90.$\pm$109   & 109$\pm$143   & 100$\pm$133   & 115$\pm$145 \\ 
$L_{\textrm{\tiny{tj}}}$ (km) & 240$\pm$314   & 221$\pm$283   & 215$\pm$271   & 233$\pm$309   & 271$\pm$360   & 206$\pm$245 \\ 
\hline
\end{tabular}
\end{center}
\end{table*}

\noindent{}{(a) \em Intensity: }Removal of stray light produces an extended 
tail in the histogram while the peak and minimum shift 
to moderately lower values similar to what was seen in Figure~\ref{umb_sta}. This 
is more evident in the histogram of peak intensity ($I_{\textrm{\tiny{max}}}$). 
The average value of $I_{\textrm{\tiny{mean}}}$ and 
$I_{\textrm{\tiny{max}}}$ are 0.24 and 0.26$I_{\textrm{\tiny{QS}}}$ 
respectively for the UN set while the same for the CR sequence are 0.24 
and 0.29$I_{\textrm{\tiny{QS}}}$ respectively. The histogram of $I_{\textrm{\tiny{mean}}}$ 
peaks at 0.21 and 0.13$I_{\textrm{\tiny{QS}}}$ for the UN and CR data
respectively while for $I_{\textrm{\tiny{max}}}$ the distribution peaks 
at 0.16 and 0.19$I_{\textrm{\tiny{QS}}}$ respectively which is in agreement
with \citet{Hamedivafa2011}. The maximum intensity of peripheral 
UDs is on an average 36\% greater than that of the central ones after stray 
light correction.

The distribution of $I_{\textrm{\tiny{max}}}/I_{\textrm{\tiny{bg}}}$ 
peaks at 1.18 and 1.58 for the UN and CR sequences respectively. In 
case of the latter the histogram exhibits a conspicuous tail with 
$I_{\textrm{\tiny{max}}}$ being $\approx$4 times greater than 
$I_{\textrm{\tiny{bg}}}$ for nearly 3\% of the UD population.
As a result, the estimated mean value of the ratio become 2.5 in 
comparison to 1.48 in the uncorrected sequence. Such a trend is similar
to the area histogram of continuum intensity at 630 nm obtained for 
numerically simulated UDs \citep{Bharti2010}.\\

\noindent{}{(b) \em Size: }Stray light removal reduces the mean effective 
UD diameter from 295 km to 272 km.
The histogram for the latter is narrower and shows a strong peak 
at 257 km while the former is more broader with a maximum around 288 km.
For the corrected sequence, central and peripheral UDs have a mean 
effective diameter of 298 km and 239 km respectively which is consistent 
with the findings of \citet{Rth2008b}.
The sizes of peripheral UDs are smaller than those of central UDs which
is similar to that of \citet{Rth2008b}.
\citet{Bharti2010} state that the
average diameter of simulated UDs is $\approx$ 320 km with the histogram
suggesting that UDs do not have a typical size. In comparison, the 
histograms of the effective diameter shown in Figure~\ref{histo} and that 
obtained by \citet{Rth2008b} tend to be nearly symmetrical.\\

\begin{figure}[!h]
\centerline{
\includegraphics[angle=90,width = 0.48\textwidth]{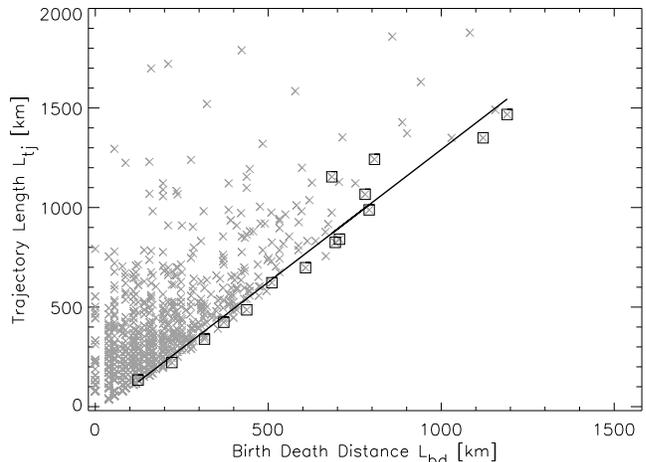}
}
\caption{Relation between $L_{\textrm{\tiny{bd}}}$ and $L_{\textrm{\tiny{tj}}}$. The scatter 
plot between birth-death distance and trajectory length shown as grey cross symbols. The 
square symbols represent maximum $L_{\textrm{\tiny{bd}}}$ for a given $L_{\textrm{\tiny{tj}}}$ 
within a bin of 100 km. The straight line is a linear fit to the scatter of the square symbols.}
\label{bd_tj}
\end{figure}

\begin{figure}[!h]
\centerline{
\includegraphics[angle=90,width = 0.56\textwidth]{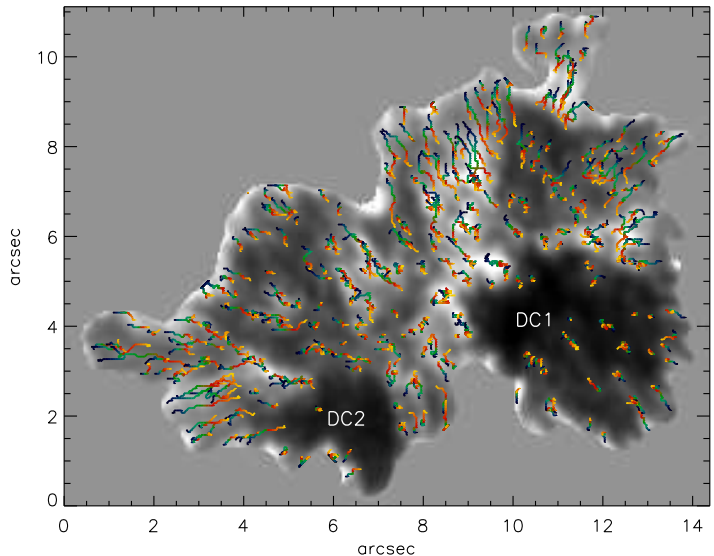}
}
\caption{Trajectories of UDs. The trajectory of UDs is overlaid on the time average image 
from the stray light corrected sequence. Blue and yellow represent birth and death respectively. 
Only UDs whose lifetime exceeds 10 min have been shown in the figure. DC1 and DC2 refer to 
the two strong dark cores in the umbra.}
\label{tracks}
\end{figure}

\noindent{}{(c) \em Horizontal Speed: }
Central as well as peripheral UDs 
tend to be mobile with speeds of $\approx$510 and 460 m~s$^{-1}$
respectively for the uncorrected set. In the corrected sequence the speeds of central and
peripheral UDs tend to be nearly the same, 460 and 440 m~s$^{-1}$, respectively,
which is in good agreement with \citet{Kilcik2011}.
\citet{Rth2008b} find peripheral UDs to be faster than central ones by 40 
m~s$^{-1}$ while \citet{Watanabe2009b} report a difference of 170 m~s$^{-1}$.
Although the above values show peripheral UDs to be slower than central UDs, 
the maximum speed of the former is nearly 200 m~s$^{-1}$ greater than the latter. In 
addition, the difference in the value is small compared to the spread in the 
distribution. The reduced speed of peripheral UDs in our case could be attributed 
to the umbral region defined by the mask described earlier.
Nearly 15\% of the population tend to be stationary in both sequences. Figure~\ref{histo} 
shows that the overall trend of the histograms for the UN and CR sets are
quite similar with some minor differences in the leading half of the distribution.\\

\noindent{}{(d) \em Lifetimes: }UDs do not exhibit a typical lifetime as is
evident from Figure~\ref{histo} (second column third row). The distribution 
is exponential with the peak coinciding with the smallest lifetime 
bin. With a cutoff of 75 s, our analysis puts
the mean value of central and peripheral UDs at 8.9 and 7.8 min respectively for 
the corrected image sequence. This is consistent with lifetimes obtained by
\citet{Hamedivafa2008} and \citet{Rth2008b}. 
According to \citet{Hamedivafa2011}, who obtained a similar exponential 
distribution, central and peripheral UDs have a typical half-life of 5 and 3 min 
respectively while \citet{Watanabe2009b} report moderately smaller lifetimes 
of 6.5 and 7.8 min respectively. These values however are much smaller than the ones
obtained by \citet{Bharti2010} who report mean values of $\approx$ 25 min for simulated UDs.
UDs which are present at the begininng or at the end of the time series 
constitute only 5\% of the total number and do not influence the lifetime histogram.\\

\noindent{}{(e) \em Trajectory: }The bottom panels of Figure~\ref{histo} correspond
to the birth-death distance and trajectory length of UDs respectively. 
The overall distribution of the above quantities appear quite similar in both image
sequences.

Nearly 55\% of the UD population have trajectory lengths of less than 150 km 
and mean horizontal speeds of 375 m~s$^{-1}$. By comparison, the rest of the 
group are relatively mobile with speeds of 550 m~s$^{-1}$.
The relation between $L_{\textrm{\tiny{tj}}}$ and $L_{\textrm{\tiny{bd}}}$
shown in Figure~\ref{bd_tj}, illustrates that for a given trajectory
length there exists a range of birth-death distances. However, the maximum 
value of $L_{\textrm{\tiny{bd}}}$ varies linearly for any $L_{\textrm{\tiny{tj}}}$
or even when the latter is averaged within a certain length range. The linear
relation is shown for the square symbols which correspond to the maximum birth-death 
distance for a given trajectory length within bins of 100 km.\\

The trajectories of UDs from their point of origin to death is shown 
in Figure~\ref{tracks}. These paths have been overlaid on the time averaged 
image from the stray light corrected sequence. The figure only shows 
the trajectories of those UDs whose lifetime exceeds 10 min. These features 
constitute 23\% of the UD population. Peripheral UDs have a tendency to move 
inward into the umbra with a nearly linear trajectory. At least two strong 
dark cores can be identified in the umbra which have been labeled DC1 and 
DC2 in the figure. One finds that UDs tend to gather or terminate 
near the edges of these dark cores which is qualitatively in agreement 
with \citet{Watanabe2009b}. Furthermore, there does not appear to be a strict
segregation of the nature of trajectories on the basis of the location/origin
of UDs. For instance, UDs with nearly linear radial tracks are not necessarily confined to 
the periphery of the umbra while those with squiggly trajectories having 
birth-death distances less than 200 km are mostly located in the inner
regions of the umbra. The latter are observed particularly in the 
bright parts of the umbra including the region between DC1 and DC2 
as well as the light bridge on the northern boundary of DC1.

\begin{table}[!h]
\begin{center}
\caption{Linear correlation coefficients (CCs) between various properties for all, central 
and peripheral UDs. The values shown in the table correspond to the uncorrected sequence 
while those in parantheses represent correlations from the stray light corrected set. CCs 
greater than 0.3 are shown in boldface.}
\label{cor}
\begin{tabular}{|c|c|c|c|}
\hline
& & &  \\
Correlation & All UDs & Central & Peripheral \\
between &  &  &  \\
\hline \hline
& & &  \\
$I_{\textrm{\tiny{max}}}$ - $V$ & -0.06(-0.004) & -0.04(0.01) & -0.01(0.01)  \\
$I_{\textrm{\tiny{max}}}$ - $L_{\textrm{\tiny{bd}}}$ & 0.24({\bf{0.31}}) & {\bf{0.35}}({\bf{0.45}}) & 0.20(0.25)  \\
$I_{\textrm{\tiny{max}}}$ - $L_{\textrm{\tiny{tj}}}$ & 0.25(0.27) & {\bf{0.30}}({\bf{0.39}}) & 0.22(0.25)  \\
$I_{\textrm{\tiny{max}}}$ - $T$ & {\bf{0.33}}(0.28) & {\bf{0.34}}({\bf{0.38}}) & {\bf{0.30}}(0.29)  \\
$D_{\textrm{\tiny{eff}}}$ - $V$ & 0.21(0.20) & 0.09(0.13) & 0.27(0.26)  \\
$D_{\textrm{\tiny{eff}}}$ - $T$ & -0.08(-0.04) & 0.27(-0.13) & -0.18(-0.06)  \\
$D_{\textrm{\tiny{eff}}}$ - $L_{\textrm{\tiny{bd}}}$ & 0.21(0.02) & 0.16(-0.16) & {\bf{0.31}}(0.16)  \\
$D_{\textrm{\tiny{eff}}}$ - $L_{\textrm{\tiny{tj}}}$ & 0.04(0.03) & 0.28(-0.10) & -0.05(0.08)  \\
$V$ - $L_{\textrm{\tiny{tj}}}$ & 0.25(0.23) & 0.23(0.17) & 0.27({\bf{0.33}})  \\
$T$ - $L_{\textrm{\tiny{tj}}}$ & {\bf{0.86}}({\bf{0.89}}) & {\bf{0.90}}({\bf{0.91}}) & {\bf{0.84}}({\bf{0.86}})  \\
$T$ - $V$ & -0.03(-0.02) & -0.01(-0.05) & -0.03(0.02)  \\
\hline
\end{tabular}
\end{center}
\end{table}

Table~\ref{cor} lists the linear correlation coefficients (CCs) between 
various properties for both uncorrected and corrected sequences 
(values within parantheses for the latter). The CCs are calculated 
between $V$, $T$, $L_{\textrm{\tiny{bd}}}$ as well as $L_{\textrm{\tiny{tj}}}$ 
with $I_{\textrm{\tiny{max}}}$ and $D_{\textrm{\tiny{eff}}}$. CCs greater 
than 0.3 are shown in boldface. The peak intensity of UDs is 
poorly correlated with its horizontal speed. If one considers the 
correlation between the peak intensity at the point of emergence with
the average velocity (total distance/total time), there is no substantial
change from the values cited in the table. Brighter, central UDs have larger 
birth-death distances as well as trajectory lengths as indicated by the 
moderate CC which is seen in both the uncorrected and corrected
sets. However, the CC worsens in the case of peripheral UDs which 
tends to bring down the same for the entire sample of UDs. 

Brighter UDs also
tend to be long lived with values of 0.38 and 0.29 for central and peripheral
UDs, respectively, as derived from the corrected sequence. A modest correlation
between size and speed is observed predominantly in peripheral UDs. In addition
larger UDs of either kind do not show any preference with lifetimes,
birth-death distances and trajectory lengths resulting from the corrected 
images. The same is the case for the uncorrected images with the exception 
of a weak correlation between size and birth-death distance for peripheral
UDs. In comparison to central UDs, peripheral UDs exhibit a better, although 
moderate, correlation between mobility and trajectory lengths. The lifetimes on the other
hand, are strongly correlated with trajectory lengths. This trend is 
expected since the trajectory length is dependent on the number of frames 
for which the feature is tracked. The lifetime of UDs is poorly correlated with
its horizontal speed.

\subsection{Spatial and Temporal Coincidence}
\label{coin}
The time sequence 
movie reveals that UDs tend to move along trajectories that are traced by 
subsequent UDs. This would suggest that there are regions/pockets in the umbra 
which favor the emergence and disappearance of UDs. There could also exist 
a time delay or a set of delays between the appearance of one UD 
and the emergence/disappearance of another at the same spatial 
location. This aspect is dealt with in the following manner. From 
the information table of each UD, one can extract the 
time interval between the i) birth of two UDs, ii) birth of one UD with the 
death of another and vice-versa and iii) death of two UDs, where all 
three classes of events refer to the same spatial location. These events can
be referred to as birth-birth (BB), birth-death-birth (BDB) and death-death (DD) 
respectively. We confine our analysis and discussion to central UDs, since
the origin of peripheral UDs is predominantly in the penumbra 
\citep{Sobotka2009a,Watanabe2009b}.

\begin{figure}[!h]
\centerline{
\vspace{15pt}
\includegraphics[angle=90,width = 0.5\textwidth]{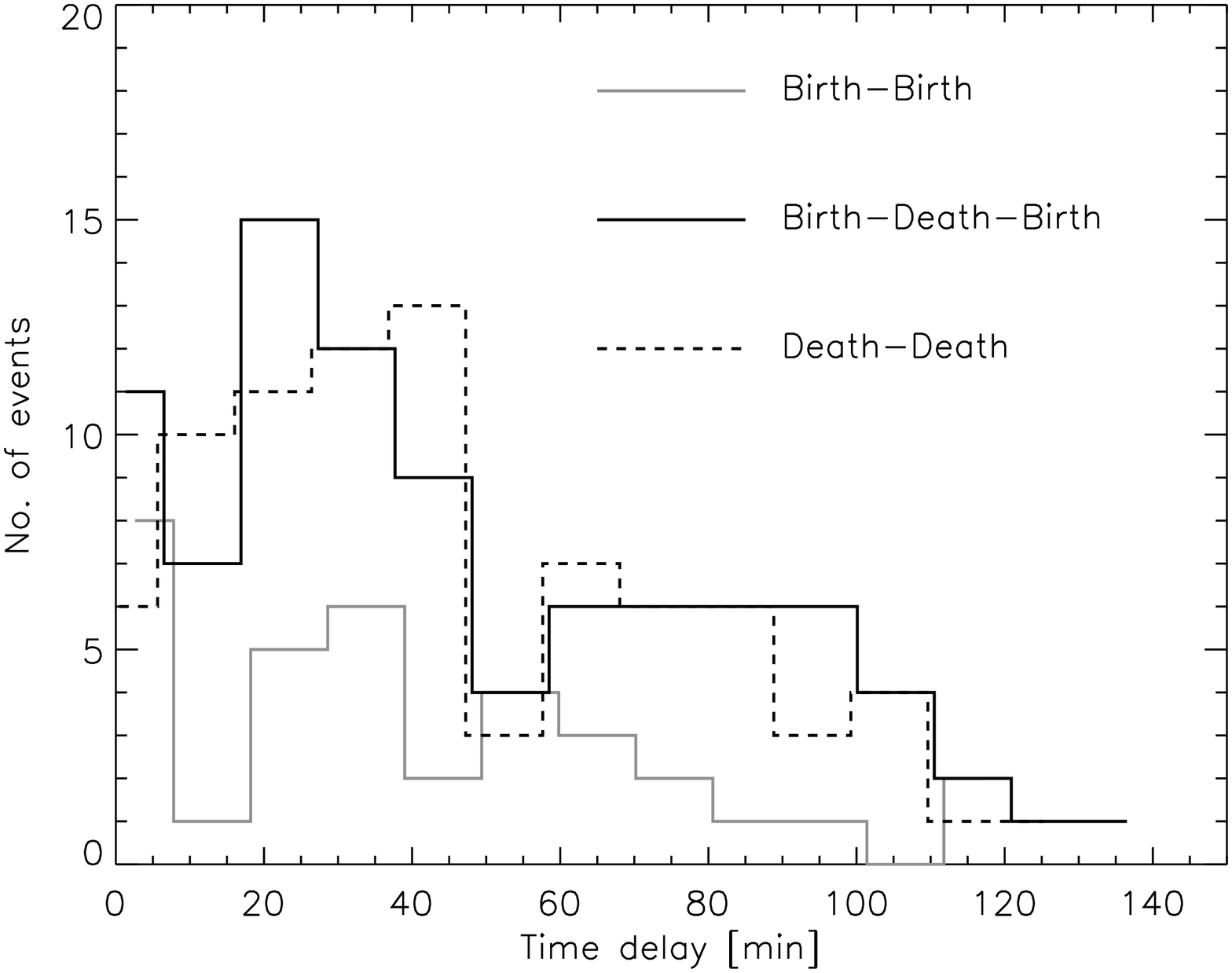}
}
\centerline{
\vspace{10pt}
\includegraphics[angle=90,width = 0.5\textwidth]{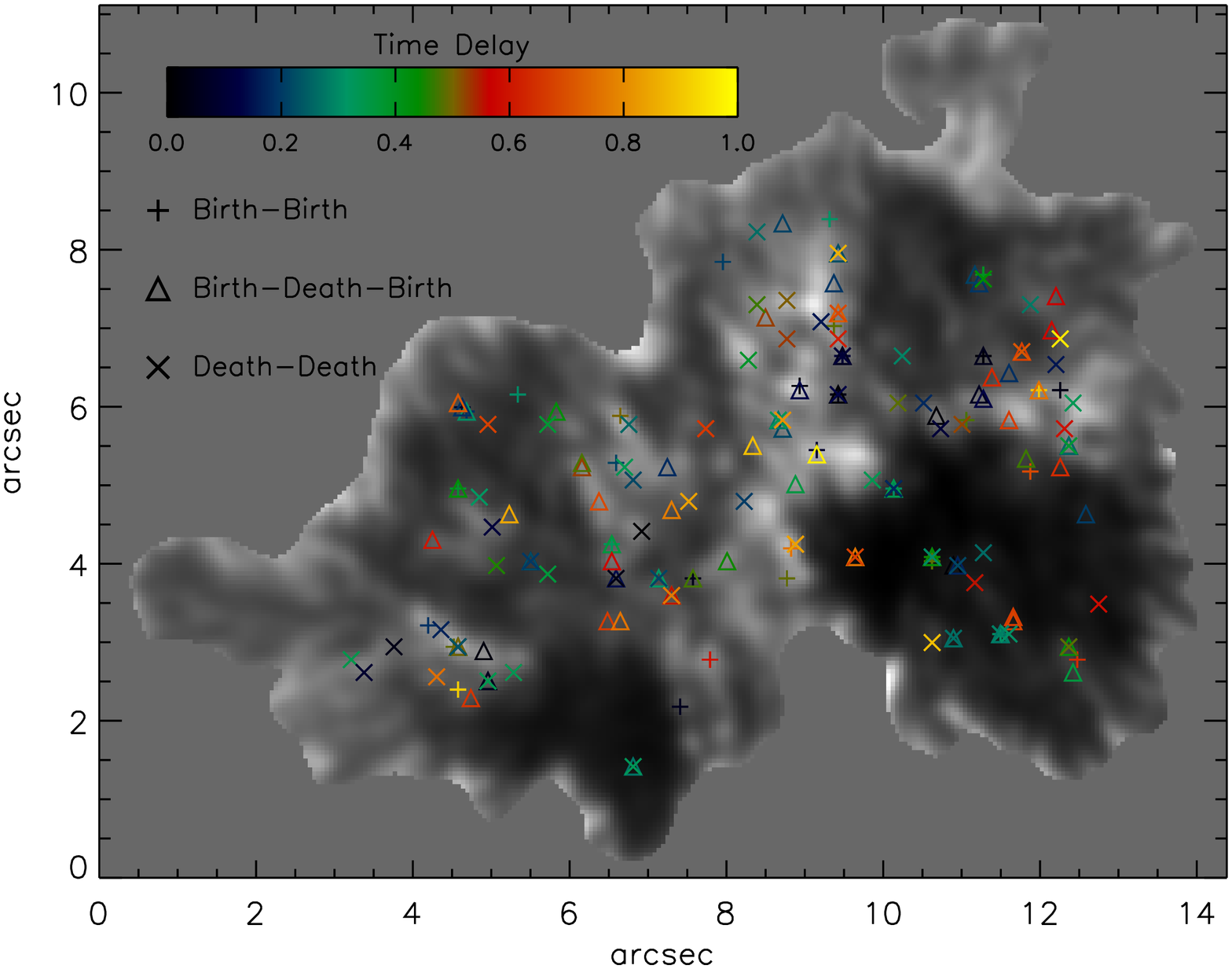}
}
\caption{{\bf{Top: }}Histograms of time delays associated with each event 
class: BB--({\it{grey}}), BDB--({\it{solid black}}), DD--({\it{black dotted}}). 
{\bf{Bottom: }}Spatial distribution of the occurrences of all three event classes 
(see text) which are indicated by the {\em plus}, {\em triangle} and {\em cross} 
symbols respectively. The color coding indicates the normalized time delay associated 
with that spatial location for each event class. BB:Birth-Birth, BDB:Birth-Death-Birth, 
DD:Death-Death.}
\label{coinc}
\end{figure}

The top panel of Figure~\ref{coinc} shows the distribution of the number of events
in the three classes respectively for all UDs whose lifetime exceeds 
150 s. 
The histograms indicate that there is no unique
time lag between successive events in either class, although BDB and DD have a
tendency to have shorter time delays than the BB class. Mean time delays of 45.5, 
52.0 and 49.8 min were estimated for the three classes respectively. If the minimum
lifetime of the UDs is varied in the range 50s - 10 min, the mean value of the distribution in
all three event classes are limited within $\pm$5 min of the above values.
Taking the average time delay for this range of lifetimes, one obtains 
46, 53 and 47 min for the each event class respectively. By comparison, the 
number of events in BB are fewer than the other two for any given minimum lifetime. 
However, for short lived UDs ($T<$150s), the number of BB and DD events are comparable 
with mean time delays of 45, 46 and 52 min in the three classes respectively.

The bottom panel of Figure~\ref{coinc} shows the spatial distribution
of the occurrences of all three event classes which are indicated by the
{\em plus}, {\em triangle} and {\em cross} symbols respectively. The color 
corresponds to the normalized time delay associated with that spatial 
location for each event class. The figure indicates that the pockets of sustained
emergence as well as disappearance are scattered everywhere in the umbra with no 
specific pattern in the distribution. The figure also shows that a particular
spatial location is not limited to the event class as at least 12 and 20 instances
were estimated to be common to BB and BDB as well as BDB and DD respectively.
While a similar characteristic of spatial coincidence of inter event classes is
seen for UDs with lifetimes of less than 2 min, the number of coincidences is 
higher namely, 19 cases each for the above two inter event classes. This 
would suggest that short lived UDs tend to reappear or emerge from the point 
of disappearance of former UDs.

\begin{table*}[!ht]
\begin{center}
\caption{Comparison of a few physical properties obtained previously from {\em Hinode}, ground-based observations and numerical simulations.}
\label{compare}
\begin{tabular}{|c|c|c|c|c|c|c|}
\hline
Reference & Source & UD Identification Algorithm &  $D_{\textrm{\tiny{eff}}}$ (km) & $I_{\textrm{\tiny{max}}}$/$I_{\textrm{\tiny{bg}}}$ & $V$ (km~s$^{-1}$)& $T$ (min) \\
\hline\hline
No Stray Light Correction & {\em Hinode}   & MLT                                 & 295$\pm$101  & 1.48$\pm$0.34 & 0.49$\pm$0.35 & 8.8$\pm$11.9  \\
\citet{Watanabe2009b}     & {\em Hinode}   & Intensity Thresholding              & 184          & 1.73          & 0.44          & 7.35          \\
With Stray Light Correction & {\em Hinode} & MLT                                 & 272$\pm$68   & 2.50$\pm$1.26 & 0.45$\pm$0.32 & 8.4$\pm$10.5  \\
\citet{Hamedivafa2008} & 0.5 m SVST        & Low-Noise Curvature Detection       & 230          & --            & $<$1.0        & 7-10 \\
\citet{Rth2008b}   & 1.0 m SST             & MLT                                 & 272$\pm$53   & 1.17$\pm$0.1  & 0.42$\pm$0.20 & 10.5$\pm$10.5 \\
\citet{Sobotka2009b}& 1.0 m SST            & Low-Noise Curvature Detection       & 125          & 2.4           & 0.34          & 9.1 \\
\citet{Kilcik2011} & 1.6 m NST             & Area \& Intensity Thresholding      & 254          & --            & 0.45          & 8.19          \\
\citet{Rempel2009a} & Simulations          & Area \& Intensity Thresholding      & 295          & --            & 0.35          & 12.9          \\
\citet{Bharti2010} & Simulations           & MLT                                 & 420          & 2.88          &    --         & 25.1          \\
\hline
\end{tabular}
\end{center}
\end{table*}

\section{Summary and Discussion}
\label{summary}
We employ a 2 hr 20 min time sequence of high resolution blue continuum 
filtergrams of the sunspot in NOAA AR 10944 from {\em Hinode} SOT 
to determine the properties of umbral dots and analyze the umbral 
fine structure after careful removal of instrumental stray light. 
The motivation of our study stems from the fact that stray 
light reduces image contrast which could influence the 
photometric and geometric properties of umbral features. After
deconvolving the filtergrams with an instrumental PSF, whose
wings describe the scattered light in the telescope, we find
that the $rms$ contrast of the images increases by a factor of 1.45.
In addition, the mean umbral intensity decreases by 
$\approx$ 17\%.

With scattered light accounted for, one is able to
identify short filamentary structures resembling penumbrae but well
separated from the umbra-penumbra boundary. The features consist
of a dark filament with bright filaments/grains adjacent to it. The 
width of the dark filament is at the resolution limit of {\em Hinode}, 
suggesting that these structures are largely unresolved. Their lifetimes
are of the order of 10-20 min during which bright grains can transit from
one filament to another remaining close to the dark filament.
The existence of similar features have already been reported by
\citet{Rimmele2008,Sobotka2009b} based on ground based observations from
larger telescopes as well as from numerical simulations \citep{Rempel2009}.
We however, do not find sub-structures in UDs namely,
the dark lanes, which represent hot convective plumes, since they are still
below the detection capability of {\em Hinode}. The observations of 
\citet{Bharti2007} are reminiscent of fragmenting light bridges that 
often exhibit dark cores along their axis \citep{Lites2004,Rimmele2008,Rohan2008}.

Various physical properties of umbral dots were determined after they
were identified in individual images and tracked in time. To that end,
a multi-level tracking algorithm was applied to both the uncorrected 
and corrected image sequences. The following properties
were analyzed : mean intensity ($I_{\textrm{\tiny{mean}}}$), 
maximum intensity ($I_{\textrm{\tiny{max}}}$), 
ratio of maximum-to-background intensity ($I_{\textrm{\tiny{max}}}/I_{\textrm{\tiny{bg}}}$),
effective diameter ($D_{\textrm{\tiny{eff}}}$), 
horizontal speed ($V$), 
lifetime ($T$), 
birth-death distance ($L_{\textrm{\tiny{bd}}}$) 
and trajectory length ($L_{\textrm{\tiny{tj}}}$). 
The mean value of the above quantities derived from the stray light corrected 
image sequence are 0.24$I_{\textrm{\tiny{QS}}}$, 0.29$I_{\textrm{\tiny{QS}}}$, 
2.5, 272 km, 0.45 km~s$^{-1}$, 8.4 min, 112 km and 221 km respectively.
The distribution of the values describing the photometric and geometric properties is more 
strongly affected by the presence of stray light while it is less severe in
the case of the kinematic parameters. 

The time sequence of corrected images was also used to determine if
umbral dots prefer to appear or disappear at specific locations in the umbra.
Furthermore, any specific time delay or a set of time delays associated with
subsequent emergence/disappearance was also determined. The events which were
sorted as birth-birth, birth-death-birth and death-death do not have a unique
time delay with a majority of events in either group having a value varying 
between 45-50 min. These values are also insensitive to the minimum 
lifetime of umbral dots. The spatial locations where these events occur 
are scattered in the umbra and in general do not highlight a specific 
pattern. In addition, there are several pixels where events from either class 
can overlap. This would suggest that umbral dots tend to reappear or emerge 
from the location of disappearance of former umbral dots.

Table~\ref{compare} gives mean 
values of a few properties obtained previously from {\em Hinode}, ground-based 
observations and numerical simulations. The values obtained from the 
uncorrected sequence are in good agreement with those reported by 
\citet{Watanabe2009b} for the same data set described in this paper, although 
they employed a different identification routine. On the other hand, the corresponding 
values obtained from the stray light corrected images are consistent with 
\citet{Hamedivafa2008,Rth2008b,Kilcik2011}. The ratio of maximum-to-background 
intensity ($I_{\textrm{\tiny{max}}}/I_{\textrm{\tiny{bg}}}$) is particularly 
in good agreement with those obtained by \citet{Sobotka2009b} and \citet{Bharti2010}.
There appears to be some differences between the results derived from 
numerical simulations by \citet{Rempel2009a} and \citet{Bharti2010} which show 
UDs to be larger, slower and long-lived than those retrieved from 
observations. The deviations between the two sets of simulations
could be attributed to the boundary conditions and treatment of numerical 
diffusivities \citep{Kilcik2011}. In the case of observations, there is 
general consensus that UDs are typically 250 km in diameter and have 
lifetimes of 10 min and horizontal speeeds of 450 m~s$^{-1}$. Another point 
of disagreement between simulations and observations is the ubiquitous presence 
of UDs in the former that do not give any indication of structuring in 
the umbra, i.e. faint light bridges and umbral dark cores. 

With the exception of intensity vs diameter, none of the
UD properties are well correlated with each other \citep{Kilcik2011}.
In simulations, the correlation coefficient is nearly 0.6 but can 
vary between 0.02 and 0.3 depending on the sample of UDs taken from relatively 
bright locations in the umbra \citep{Kilcik2011}. The 
monotonicity seen in simulations, exists with a significant scatter 
between the intensity and diameter of UDs. The differences between observations and 
simulations, needs to be addressed with complementary polarimetric observations at high 
spatial resolution ($<$0\farcs15) which could provide additional constraints for 
MHD models. This is necessary for determining the physical mechanisms driving
UDs and their kinematics, for which instruments such as the CRISP \citep{Scharmer2008} 
and the GFPI \citep[Gregor Fabry-P$\acute{\textrm{e}}$rot Interferometer;][]{Denker2010}
will be crucial.

\acknowledgments
Our sincere thanks to the {\em Hinode} team for providing the data
used in this paper. Hinode is a Japanese mission developed and
launched by ISAS/JAXA, with NAOJ as domestic partner and NASA and STFC
(UK) as international partners. It is operated by these agencies in
co-operation with ESA and NSC (Norway). The authors thankfully acknowledge
the comments and suggestions of the referee. Our work has been partially 
funded by the Spanish MICINN through projects AYA2011-29833-C06-04 and 
PCI2006-A7-0624, and by Junta de Andaluc\' {\i}a through 
project P07-TEP-2687, including a percentage from European
FEDER funds.

\end{document}